\documentclass[aps,prb,twocolumn,superscriptaddress,showpacs]{revtex4}
\usepackage{graphicx}
\usepackage{dcolumn}
\usepackage{bm}
\usepackage{amsmath}

\begin{document}

\title{
X-ray Anomalous Scattering of Diluted Magnetic Oxide Semiconductors: Possible Evidence of Lattice Deformation for High 
Temperature Ferromagnetism
}


\author{Takeshi Matsumura}
\email[]{tmatsu@iiyo.phys.tohoku.ac.jp}
\author{Daisuke Okuyama}
\author{Shinya Niioka}
\author{Hideaki Ishida}
\author{Tadashi Satoh}
\author{Youichi Murakami}
\affiliation{Department of Physics, Graduate School of Science, Tohoku University, Sendai, 980-8578, Japan}

\author{Hidemi Toyosaki}
\author{Yasuhiro Yamada}
\author{Tomoteru Fukumura}
\affiliation{Institute for Materials Research, Tohoku University, Sendai, 980-8577, Japan}
\author{Masashi Kawasaki}
\affiliation{Institute for Materials Research, Tohoku University, Sendai, 980-8577, Japan}
\affiliation{CREST, Japan Science and Technology Agency, Tokyo 102-0075, Japan}


\date{\today}

\begin{abstract}
We have examined whether the Co ions crystallographically substitute on the Ti sites in rutile and anatase Ti$_{1-x}$Co$_{x}$O$_{2-\delta}$  
thin films that exhibit room-temperature ferromagnetism. Intensities of the x-ray Bragg reflection from the films 
were measured around the $K$-absorption-edge of Co. If the Co ions randomly substitute on the Ti sites, the intensity should exhibit 
an anomaly due to the anomalous dispersion of the atomic scattering factor of Co. However, none of the anatase and rutile samples 
did exhibit an anomaly, unambiguously showing that the Co ions in Ti$_{1-x}$Co$_{x}$O$_{2-\delta}$ are not exactly located 
at the Ti sites of TiO$_2$. The absence of the anomaly is probably caused by a significant deformation of the local structure around 
Co due to the oxygen vacancy. We have applied the same method to paramagnetic Zn$_{1-x}$Co$_{x}$O thin films and obtained direct 
evidence that the Co ions are indeed substituted on the Zn sites. 
\end{abstract}


\maketitle


\section{Introduction}
Diluted magnetic semiconductors (DMS) have been attracting considerable attention as a promising candidate material 
for spintronics devices. The basic concept is to utilize spin polarized carriers created by strong exchange interaction 
between localized $d$-electrons of diluted magnetic ions and itinerant $sp$-band electrons.\cite{Dietl00,Ohno98} 
Since the discovery of room temperature ferromagnetism in Co-doped anatase and rutile TiO$_2$ thin film,\cite{Matsumoto01a,Matsumoto01b} 
there have been a number of reports on high-$T_C$ ferromagnetism observed in various kinds of 
DMS, particularly in oxides.\cite{Fukumura05,Janisch05} 
Co-doped TiO$_2$ is the material that has been studied most intensively. 
However, the origin of the ferromagnetism has not yet been elucidated. One of the experimental 
issues is whether the ferromagnetism indeed originates from the Co spins that are randomly substituted on the Ti sites, 
and not from the segregated Co clusters. There have also been a number of reports that ascribe the ferromagnetism to 
the Co segregation.\cite{Kim03,Shinde04,Higgins04}  

Strong evidence in support of intrinsic ferromagnetism has been given by the observations of 
anomalous Hall effect (AHE) and magnetic circular dichroism (MCD).\cite{Toyosaki04,Toyosaki05a,Fukumura03,Yamada04,Ueno07} 
If the ferromagnetism were intrinsic, the charge carriers should be spin polarized due to the exchange interaction 
with the Co spins. AHE and MCD are considered to probe the ferromagnetic response of the carriers introduced by the oxygen deficiency. 
In addition, rutile Ti$_{1-x}$Co$_{x}$O$_{2-\delta}$ has already been functionalized as a spin tunneling junction 
working up to 180 K.\cite{Toyosaki05b} 
Other studies that focus on this issue are mainly based on spectroscopic measurements 
such as x-ray absorption fine structure (XAFS) for samples in which no Co segregation was recognized by imaging probes 
such as scanning electron microscope (SEM), atomic force microscope (AFM), and transmission electron microscopy 
(TEM).\cite{Chambers03,Shimizu04,Griffin05,Murakami04} 
In these spectroscopic measurements, the spectral line shapes differ from that of Co metal. 
They state that the local structure of the Co site is close to those of Co oxides such as CoTiO$_3$, in which the Co ion is surrounded by 
six oxygen atoms, forming a CoO$_6$ cluster, and that the valence state of Co is divalent.  
Cui \textit{et al.} performed electron beam diffraction and composition analysis for anatase TiO$_2$ in a selected area 
without Co segregation and confirmed both the diffraction peaks of TiO$_2$ and the existence of Co.\cite{Cui04} 
These results seem to support that the Co ions randomly substitute on the Ti sites. 

Although these spectroscopic methods provide valuable information on the local structure around Co, however, they give little information 
on the crystallographic position of Co and the orientation of the CoO$_6$ cluster: the former could be interstitial and the latter 
could be deformed due to the oxygen vacancy. 
In this context, we still do not have unambiguous evidence that Co indeed substitutes for Ti in a crystallographic sense. 
The fact that Co is not soluble in TiO$_2$ in a thermodynamically stable manner also casts doubt on the assumption 
that Co substitutes for Ti.\cite{Li03} 
From a theoretical point of view, knowledge on the local structure of a Co ion is of fundamental importance to construct a valid model and proceed the calculations on the electronic states of $3d$ and $sp$-band electrons. 

In the present paper, we report on the results of x-ray diffraction utilizing anomalous scattering from Co.  
This is a fundamentally different approach than spectroscopy. 
We observe a Bragg peak as a result of the interference among the x rays scattered from many Co ions in the sample and discuss the 
average crystallographic site of Co in the TiO$_2$ lattice. 
If Co substitute exactly on the Ti site, the Co ions have the same periodicity as the TiO$_2$ lattice and contribute to the Bragg peak. 
Our results on anatase and rutile Co-doped TiO$_2$ films, however, lead us to a conclusion 
that the Co ions are not exactly located on the Ti sites, implying a significant lattice deformation. 
On the other hand, it is shown that Co indeed substitutes on the Zn sites in paramagnetic Co-doped ZnO thin films. 

\section{Experiment}
The basic principle of the method is simple and direct. 
If the Co ions of concentration $x$ randomly substitute on the Ti sites of TiO$_2$, the unit-cell structure-factor can be expressed as 
\begin{eqnarray}
F &=& \sum_{i} \{(1-x)f_{\text{Ti}} + xf_{\text{Co}}\}\exp i\bm{\kappa}\cdot\bm{R}_i^{\text{(Ti)}} \nonumber \\
  && + \sum_{i} f_{\text{O}} \exp i\bm{\kappa}\cdot\bm{R}_i^{\text{(O)}}\;,   \label{eq:1}
\end{eqnarray}
where $f_{\text{Ti}}$, $f_{\text{Co}}$, and  $f_{\text{O}}$ are the atomic scattering factors of Ti, Co, and O, respectively. 
$\bm{R}_i^{\text{(Ti)}}$ and $\bm{R}_i^{\text{(O)}}$ represent the $i$-th atomic site of Ti and O in the unit cell, respectively. 
$\bm{\kappa}$ is the scattering vector. Here, the atomic scattering factors are energy dependent and are 
generally expressed as 
\begin{equation}
f(E) = f^{0} + f'(E) + if''(E)\;,
\end{equation}
where $f^0$ is the Thomson scattering factor and $f'$ and $f''$ are the real and imaginary parts, respectively, 
of the anomalous scattering factor, which exhibit a significant anomaly near an absorption edge of the element. 
Therefore, when we measure the energy dependence of the intensity of a Bragg reflection from the Co-doped TiO$_2$ film, 
the intensity should exhibit an anomaly at the absorption edge of Co if the structure factor involves $f_{\text{Co}}$. 
This measurement can be performed using a synchrotron radiation source, where the incident energy of the x rays can be varied. 
Information of the crystal structures, reflection indices examined in the present experiment, and their structure factors 
are listed in Table~\ref{table:1}. 
In rutile, the Ti atoms occupy the crystallographic site of $2a$: (0, 0, 0) and ($\frac{1}{2}$, $\frac{1}{2}$, $\frac{1}{2}$). 
In anatase, they are at the $4a$ site: (0, 0, 0), (0, $\frac{1}{2}$, $\frac{1}{4}$), ($\frac{1}{2}$, $\frac{1}{2}$, $\frac{1}{2}$), and 
($\frac{1}{2}$, 0, $\frac{3}{4}$). 
The Zn atoms of ZnO occupy the $2b$ site of the wurtzite structure: 
($\frac{1}{3}$, $\frac{2}{3}$, 0) and ($\frac{2}{3}$, $\frac{1}{3}$, $\frac{1}{2}$). 
\begin{table*}
\caption{Reflection index and the structure factor examined in this experiment. }
\label{table:1}
\begin{ruledtabular}
\begin{tabular}{lllll}
sample & structure & space group & index & structure factor \\
\hline 
Ti$_{1-x}$Co$_{x}$O$_2$ & anatase & $I4_1/amd$ (\#141) & 0 0 4 & $F=4(1-x)f_{\text{Ti}}+4xf_{\text{Co}}+8f_{\text{O}}\cos 1.66\pi$ \\
Ti$_{1-x}$Co$_{x}$O$_2$ & rutile & $P4_2/mnm$ (\#136) & 2 0 2 & $F=2(1-x)f_{\text{Ti}}+2xf_{\text{Co}}+4f_{\text{O}}\cos 1.22\pi$ \\
Zn$_{1-x}$Co$_{x}$O &  wurtzite & $P6_3mc$ (\#186) & 0 0 2 & $F=2(1-x)f_{\text{Zn}}+2xf_{\text{Co}}+2f_{\text{O}}(\cos 1.53\pi + i\sin 1.53\pi)$
\end{tabular}
\end{ruledtabular}
\end{table*}

We used the same rutile Ti$_{1-x}$Co$_{x}$O$_{2-\delta}$ epitaxial thin film samples as those studied in 
Refs.~\onlinecite{Toyosaki04} and \onlinecite{Toyosaki05a}. 
The films were deposited on TiO$_2$ buffered r-sapphire substrates by laser molecular beam epitaxy (MBE). 
Oxygen deficiency $\delta$ is controlled by the oxygen partial pressure varying from 10$^{-4}$ to 10$^{-8}$ Torr. 
A systematic relationship among Co concentration $x$, oxygen partial pressure, carrier density, conductivity, 
ferromagnetic moment, AHE, and MCD is well established for these samples. 
Results of x-ray photoemission spectroscopy (XPS) and XMCD at the Co $L_{2,3}$-edges for these samples are also 
reported,\cite{Quilty06,Mamiya06} both of which conclude that the spectrum is that of a high-spin Co$^{2+}$ ion in a 
crystal field of oxygen octahedron and that the room-temperature ferromagnetism is intrinsic. 
Anatase Ti$_{1-x}$Co$_{x}$O$_{2-\delta}$ epitaxial thin film was deposited on LaAlO$_3$-(001) substrate by 
laser MBE in an oxygen partial pressure of $1\times10^{-6}$ Torr and at a growth temperature of 700 $^{\circ}$C. 
The appearance of room-temperature ferromagnetism was checked by a SQUID magnetometer and no Co segregation was 
recognized by AFM and SEM. Zn$_{1-x}$Co$_{x}$O epitaxial thin films are the same as those of 
Ref.~\onlinecite{Jin01}. These exhibit large MCD without an appearance of ferromagnetism, although several studies 
reported high-$T_C$ ferromagnetism in the same compound.\cite{Fukumura05,Janisch05} 

X-ray diffraction experiments were performed using four-circle diffractometers installed at beamlines 1A, 4C, and 16A2 of the 
Photon Factory in KEK, Japan. The incident beam was monochromatized by a Si-111 double crystals and focused by 
a bent cylindrical mirror. 
The energy was calibrated using the absorption edge of a Co metal foil. 
For each thin-film sample, we first measured the fluorescence spectrum near 
the Co $K$-edge to check if Co was actually included in the area where the beam was irradiated. Next, we measured 
the energy dependence of the intensity of the Bragg reflection. The typical beam size was $\sim 1\times1$ mm$^2$. 
All the measurements were carried out at room temperature. 

\section{Experimental Results}
\begin{figure}[tb]
\begin{center}
\includegraphics[width=7.5cm]{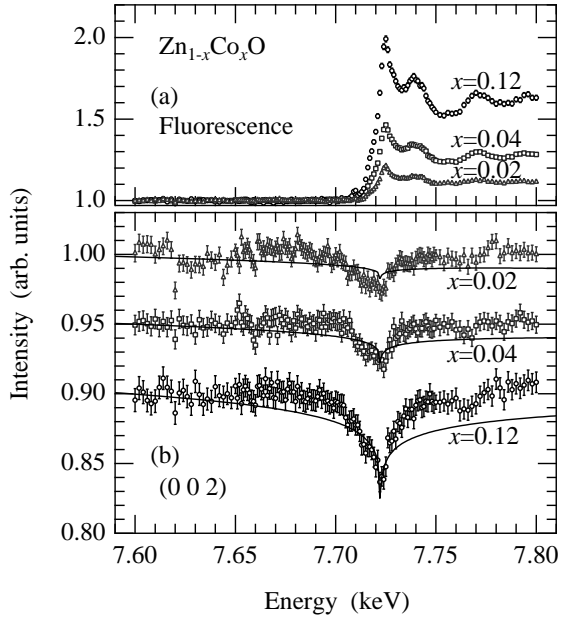}
\end{center}
\caption{(a) Fluorescence spectra of Zn$_{1-x}$Co$_{x}$O as a function of Co concentration $x$. 
(b) X-ray energy dependences of the intensity of the 002 Bragg reflection. Data are shifted for $x$=0.04 and 0.12. 
Lines represent the calculated curves. 
}
\label{fig1}
\end{figure}

The results for Zn$_{1-x}$Co$_{x}$O with $x=0.02$, 0.04, and 0.12 are shown in Fig.~\ref{fig1}. 
The base line of each spectrum is normalized to unity. The step in the fluorescence at 7.72 keV, the $K$-edge of Co, is 
roughly proportional to the Co concentration, indicating that Co is actually included in the irradiated area with concentrations 
proportional to the nominal value.  
Energy dependence of the intensity of the 002 Bragg reflection also exhibits a clear anomaly at the $K$-edge. 
This directly indicates that the Co ions indeed substitute on the Zn sites randomly. Comparison with the calculated curve represented 
by the lines is also satisfactory.\cite{SCM-AXS} 

\begin{figure}[tb]
\begin{center}
\includegraphics[width=7.5cm]{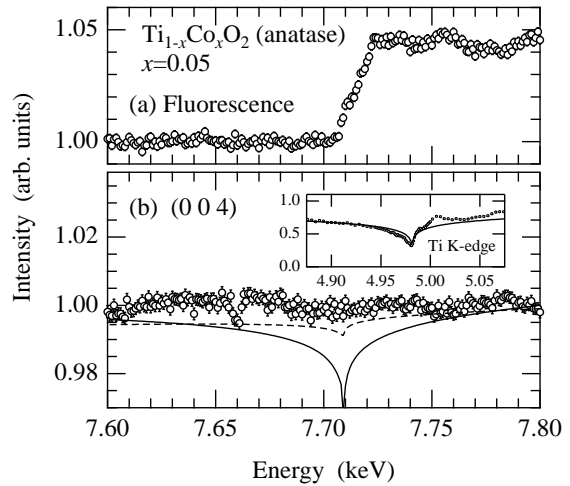}
\end{center}
\caption{(a) Fluorescence spectrum of anatase Ti$_{1-x}$Co$_{x}$O$_{2-\delta}$ for $x=0.05$. 
(b) X-ray energy dependence of the intensity of the 004 Bragg reflection.  Solid line represents a calculated curve assuming 
random substitution of Ti with 5\% Co ions. Dashed line represents a simulation considering local deformations as described in the text. 
Inset shows the result around the $K$-edge of Ti.
}
\label{fig2}
\end{figure}

Ti$_{1-x}$Co$_{x}$O$_{2-\delta}$ samples exhibit contrasting results with Zn$_{1-x}$Co$_{x}$O as described in the following. 
Figure~\ref{fig2} shows the results for anatase Ti$_{1-x}$Co$_{x}$O$_{2-\delta}$ with a nominal concentration of $x$=0.05. 
The base lines are normalized to unity. The fluorescence spectrum indicates that the Co ions are indeed included in the sample. 
However, the intensity of the 004 Bragg reflection does not exhibit any anomaly at the absorption edge of Co. 
If the Co ions with $x=0.05$ substitute for Ti, an anomaly as demonstrated by the solid line is expected, 
which is as large as about 5\% of the anomaly actually observed at the $K$-edge of Ti as shown in the inset. 
These results mean that the doped Co ions are not located exactly on the Ti sites, although the Co ions indeed exist in the sample.  

\begin{figure}[tb]
\begin{center}
\includegraphics[width=7.5cm]{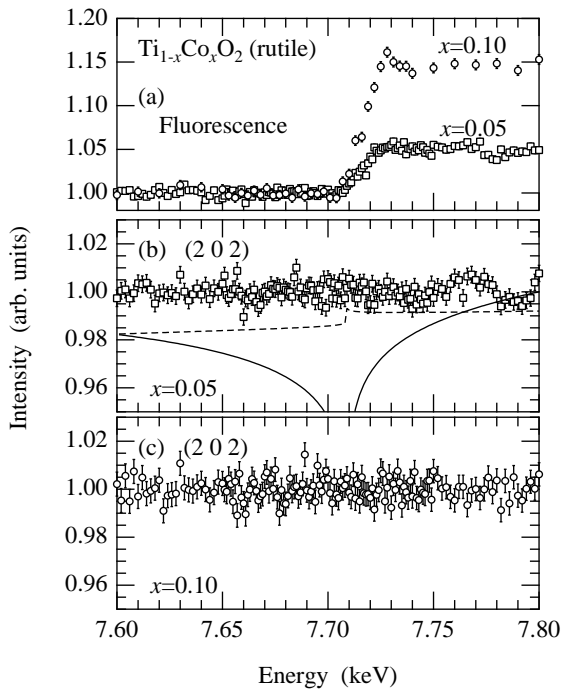}
\end{center}
\caption{(a) Fluorescence spectra of rutile Ti$_{1-x}$Co$_{x}$O$_{2-\delta}$ for $x=0.05$ and $x=0.1$. 
(b),(c) X-ray energy dependence of the intensity of the 202 Bragg reflection for $x=0.05$ and $x=0.1$. 
Solid line represents a calculated curve assuming random substitution of Ti with 5\% Co ions. Dashed line represents a 
simulation considering local deformations as described in the text.
}
\label{fig3}
\end{figure}

Figure~\ref{fig3} shows the results for rutile Ti$_{1-x}$Co$_{x}$O$_{2-\delta}$ with nominal concentrations of $x=0.05$ and 0.1, 
grown under an oxygen partial pressure of 10$^{-7}$ Torr. 
The fluorescence spectra show that the Co ions indeed exist in the samples with actual concentrations proportional to the nominal value. 
However, as in anatase, the intensities of the 202 Bragg reflection do not exhibit any anomaly at the absorption edge, 
even in the high concentration sample of $x=0.1$.  
These results again mean that the doped Co ions are not exactly on the Ti sites, although the Co ions indeed exist in the film. 
Since we expect an anomaly as large as the one demonstrated by the solid line in Fig.~\ref{fig3}, the actual amount of substitution, 
if any, is estimated to be much less than 1 \% both for $x=0.05$ and 0.1. 
The measurements on other reflections, e.g., 103 for anatase and 101 for rutile, and also on a few other samples, 
did not exhibit any anomaly. 

\section{Discussions}
\begin{figure}[t]
\begin{center}
\includegraphics[width=7.5cm]{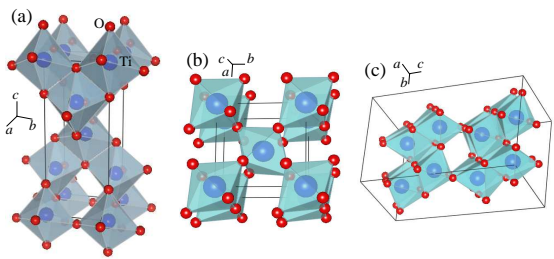}
\end{center}
\caption{(Color online) Crystal structure of (a) anatase-TiO$_2$, (b) rutile-TiO$_2$, and 
(c) Ti$_4$O$_7$ Magneli phase.\cite{VICS-II} 
}
\label{fig4}
\end{figure}

The present experimental results unambiguously show that the atomic scattering factor of Co is not included in the 
structure factor of either anatase or rutile Ti$_{1-x}$Co$_{x}$O$_{2-\delta}$. 
In other words, the Co ions, on average, do not occupy the Ti site nor any specific crystallographic site in the TiO$_2$ unit cell;  
therefore, the interference among x-rays scattered from randomly distributed Co ions are prevented. 
In contrast, from spectroscopic measurements such as XAFS, XPS and XMCD, it is concluded that the local environment of Co is 
close to that of oxygen octahedron.\cite{Chambers03,Shimizu04,Griffin05,Murakami04,Cui04,Quilty06,Mamiya06} 
In addition, the strong correlation among Co concentration $x$, oxygen deficiency $\delta$, conductivity, ferromagnetism, AHE, and MCD investigated in 
Refs.~\onlinecite{Toyosaki04,Toyosaki05a,Fukumura03,Yamada04,Ueno07} support that the carriers are associated with the ferromagnetism originating 
from the randomly distributed Co ions. 
Taking all these experimental results into consideration, we speculate that the doped Co ions exist in a locally deformed structure, 
although they are randomly distributed in the sample without making a Co metal-cluster. 
When a Co$^{2+}$ ion is substituted for a Ti$^{4+}$ in TiO$_2$, an oxygen vacancy is necessarily created 
to maintain the charge neutrality. As a result, the number of oxygen in the ligands becomes less than six,\cite{Chambers03} 
which would lead to a deformation of the local structure around Co and a deviation of Co from the exact Ti site, i.e., the 
$2a$ site in rutile and the $4a$ site in anatase.  
There is also a possibility that Co substitutes for the interstitial sites among oxygen octahedrons. 
In rutile TiO$_2$, in particular, the interstitial occupation at positions such as ($\frac{1}{2}$, 0, 0), (0, $\frac{1}{2}$, 0), ($\frac{1}{2}$, 0, $\frac{1}{2}$), 
and (0, $\frac{1}{2}$, $\frac{1}{2}$) leads to a structure similar to the Magneli phase as illustrated in Fig.~\ref{fig4}, 
from which we may speculate that the interstitial site could be also stable for Co. 

In order to examine if such a deformation could suppress the $K$-edge anomaly of Co, we have made simulations assuming 
random shifts of the Co ions from the Ti site or the interstitial site. The result is demonstrated by the dashed lines in Figs.~\ref{fig2} and \ref{fig3}.  
The process of the simulation is as follows. 
First, 5\% of Co is doped into a super cell of $10\times 10\times 10$ unit cells, randomly 
substituting for the Ti sites in anatase and the interstitial sites in rutile, respectively.\cite{interstitial} 
At this stage, the anomaly as demonstrated by the solid lines in Fig.~\ref{fig2} and \ref{fig3} appears because the Co ions occupy a 
specific crystallographic site and Eq.~(\ref{eq:1}) is valid.  This is the first kind of randomness. 
Next, we consider the second kind of randomness; i.e., one oxygen vacancy is randomly created in the local octahedron at the Co site,  
and the Co atom is shifted to the oxygen vacancy by 1~\AA\ in anatase and 0.6~\AA\ in rutile, respectively.  
Then, we calculate $|F|^2$ for the super cell, which is shown by the dashed lines in the figures. 
Although it is not our intention to emphasize that this is really the case, the second kind of randomness explains that the anomaly at the edge 
becomes very weak; absence of the anomaly in the 103 reflection of anatase and 101 of rutile can also be explained. 
In this simulation, the weakened anomaly is associated with the increased number of possible Co sites by a factor six as a result of the 
second kind of randomness. 
There should also be a tilting of the octahedron due to the vacancy, which would weaken the anomaly even more because 
it further increases the possible positions and weakens the correlation among the Co sites; the relatively large shift values assumed above 
can be reduced. 
The spectroscopic measurements, which analyze the local structural environment on average, may hide information on this kind of 
deformation we considered here. 
In the case of Zn$_{1-x}$Co$_{x}$O, on the other hand, oxygen vacancy is not necessary when doping because both Zn and Co are divalent; 
then, there arises little deformation in the local structure and the doped Co sits exactly on the Zn site, which is the prerequisite for Eq.~(\ref{eq:1}). 

To search for possible deformed structure or another phase that could contain Co such as the Magneli phase, 
we performed additional x-ray diffraction experiment using an imaging plate Debye-Scherrer camera 
installed at beamline 1B. However, no other phase was detected as a diffraction peak 
with its intensity higher than $\sim 0.2$\% of the strongest peak of rutile or anatase structure of the TiO$_2$ film. 
Therefore, the deformed structure, if it existed, is suspected to be short ranged and randomly oriented, 
which only gives rise to incoherent scattering. 

Although the position and the local structure of Co is still uncertain, our experimental results support a theoretical investigation 
that the oxygen vacancy near Co induces structural deformation and enhances the spin density 
associated with the ferromagnetism.\cite{Weng04} 
A recent theoretical model for high-$T_C$ ferromagnetism in oxide DMS is also based on the oxygen vacancy, 
which causes impurity band exchange.\cite{Coey05} 
Elucidation of the microscopic structure around Co and its relation with the mechanism of the ferromagnetism is strongly required. 
Structural analysis by x-ray fluorescence holography, which determines the three dimensional atomic arrangement around the fluorescing atom, 
could solve this problem.\cite{Hosokawa07} 

\section{Conclusion}
By utilizing x-ray anomalous dispersion, we have directly examined whether the doped Co ions substitute for Ti 
in anatase and rutile Ti$_{1-x}$Co$_{x}$O$_{2-\delta}$ for well characterized thin-film samples exhibiting intrinsic high-$T_C$ ferromagnetism. 
Although the intensity of the Bragg reflections should exhibit an anomaly at the $K$-edge of Co if the Co ions were randomly substituted 
exactly on the Ti site, no anomaly was detected in the experiment, indicating that the Co ions are not located exactly on the Ti site. 
However, the fluorescence spectra show that the Co ions exist in the sample in a certain form;  
XPS and XMCD spectra on the identical sample support that the Co ions are randomly distributed and are surrounded by the oxygens. 
These contrasting results suggest that the local structure around Co is strongly deformed, leading to a significant shift of Co from 
the high symmetry positions of Ti sites or interstitials, probably because of the oxygen vacancy. 
We have proposed a scenario how the anomaly disappears by the local deformations. 
On the other hand, in our paramagnetic Zn$_{1-x}$Co$_{x}$O thin film, the substitution of Co for the Zn sites has been verified. 
These results may imply a significant role of lattice deformation for the high-$T_C$ ferromagnetism in oxide DMS's. 

\begin{acknowledgments}
We wish to acknowledge the technical support of Y. Wakabayashi and H. Sawa during the experiments at the photon factory. 
This work was supported by a Grant-in-Aid for Scientific Research on Priority Area, "Invention of anomalous quantum materials", 
from the Ministry of Education, Science, Sports and Culture of Japan. 
T. F is supported by NEDO, Industrial Research Grant Program (05A24020d). 
\end{acknowledgments}


\begin{thebibliography}{99}
\bibitem{interstitial} If Co atoms were substituted for the Ti site in rutile, disappearance of the anomaly both in 202 and 101 reflections cannot be explained simultaneously in the present model calculation.
\end{thebibliography}

\end{document}